# A Giant Crater on 90 Antiope?

P. Descamps[1], F. Marchis[1,2,3], T. Michalowski[4], J. Berthier[1], J. Pollock[5], P.Wiggins[6], M. Birlan[1], F. Colas[1], F. Vachier[1], S. Fauvaud[7], M. Fauvaud[7], J.-P. Sareyan[8], F. Pilcher[9], D.A. Klinglesmith[10]

[1] Institut de Mécanique Céleste et de Calcul des Éphémérides, Observatoire de Paris, UMR 8028 CNRS, 77 av. Denfert-Rochereau 75014 Paris, France
[2] University of California at Berkeley, Department of Astronomy, 601 Campbell Hall, Berkeley, CA 94720, USA
[3] SETI Institute, 515 N. Whisman Road, Mountain View CA 94043, USA
[4] Astronomical Observatory, Adam Mickiewicz University, Sloneczna 36, 60-286 Poznan, Poland
[5] Appalachian State University, Department of Physics and Astronomy, 231 CAP Building, Boone, NC 28608, USA
[6] Tooele Utah 84074, USA
[7] Observatoire du Bois de Bardon, 16110 Taponnat, France
[8] Observatoire de la Côte d'Azur, BP 4229, 06304 Nice cedex 4, France
[9] Organ Mesa Observatory, 4438 Organ Mesa Loop, Las Cruces, NM 88011 USA
[10] Magdalena Ridge Observatory, New Mexico Tech, Socorro, NM 87801, USA

Pages: 30

Tables: 3

Figures: 8

*Corresponding author:*
**Pascal Descamps**
IMCCE, Paris Observatory
77, avenue Denfert-Rochereau
75014 Paris
France

descamps@imcce.fr
Phone: 33 (0)140512268
Fax: 33 (0)146332834

**Abstract**

Mutual event observations between the two components of 90 Antiope were carried out in 2007-2008. The pole position was refined to $\lambda_0$ = 199.5±0.5° and $\beta_0$ = 39.8±5° in J2000 ecliptic coordinates, leaving intact the physical solution for the components, assimilated to two perfect Roche ellipsoids, and derived after the 2005 mutual event season (Descamps et al., 2007). Furthermore, a large-scale geological depression, located on one of the components, was introduced to better match the observed lightcurves. This vast geological feature of about 68 km in diameter, which could be postulated as a bowl-shaped impact crater, is indeed responsible of the photometric asymmetries seen on the "shoulders" of the lightcurves. The bulk density was then recomputed to 1.28 ±0.04 $gcm^{-3}$ to take into account this large-scale non-convexity. This giant crater could be the aftermath of a tremendous collision of a 100-km sized proto-Antiope with another Themis family member. This statement is supported by the fact that Antiope is sufficiently porous (~50%) to survive such an impact without being wholly destroyed. This violent shock would have then imparted enough angular momentum for fissioning of proto-Antiope into two equisized bodies. We calculated that the impactor must have a diameter greater than ~17 km, for an impact velocity ranging between 1 and 4 km/s. With such a projectile, this event has a substantial 50 % probability to have occurred over the age of the Themis family.

## 1. Introduction

The asteroid 90 Antiope is the first doubly synchronous asteroid discovered by direct imaging in the main belt of asteroids (Merline et al., 2000). Antiope's two components, 91 and 86 km in diameter in average, circle each other every 16.505 hours at a distance of 171 kilometers (Descamps et al., 2007). With a subtended angular size of 0.016 arcseconds, they are too small to be resolved individually by largest telescopes equipped with Adaptive Optics systems whose spatial resolution is of 0.035 arcseconds at best (on a 10-m telescope in K band). On the other hand they can be readily separated at elongation (separation of 0.06 arcseconds).

Nonetheless, a recent detailed study based on lightcurve morphology analysis of a large photometric survey, carried out during 2005 mutual events (Descamps et al., 2007), has highlighted that both components have shapes very close to the ones predicted for rotating, tidally locked, fluid bodies, according to the solutions of the problem first formulated by Roche in 1849. Their shapes are surprisingly nearly spherical, just slightly squashed into ellipsoids. The observed eclipses and occultations, occurring as the components transit each other as seen from Earth, have enabled us to produce reliable physical and orbital parameters for the system (Table 2).

However, the Roche ellipsoids model was not able to reproduce small photometric asymmetries appearing on the "shoulders" of the observed lightcurves. Fig. 1 shows these post-eclipse shallow concavities on the composite rotational lightcurve obtained with the 1-meter telescope at SAAO on June 2005. These features are denoted as "anomalies" because they appear as deviations from the smooth, symmetrical lightcurve (dashed curve on Fig. 1) that results from an event involving two perfect Roche ellipsoids. These discrepancies were seen on all lightcurves during the 2005 mutual events, confirming its physical authenticity. Quite evidently, something is causing the system to fade by almost 0.04 mag, during a little more than one hour, twice over the course of a full rotation. What might the origin be?

In this paper we present new observations of mutual events observed from November 2007 to March 2008 and an improved model that fits the observed lightcurves of Antiope including the asymmetries.

## 2. The 2007-2008 mutual events campaign

### 2.1 Observations

In late 2007 – early 2008, the binary system of 90 Antiope experienced a new series of mutual events due to its edge-on configuration as seen from the Earth. Mutual events, lasting about three hours, occurred at 8.25h intervals over three months around the opposition of early January 2008. Antiope was a V=13.2 magnitude object with J2000 equatorial coordinates $\alpha$ = 00h28mn and $\delta$ = +24°25' at opposition, on 2008 January 5. We undertook differential photometric observations through an amateur-professional collaborative network with small telescopes up to 1.2 m in aperture diameter and equipped with a visible CCD detector. Table 1 gives the list of observers along with the main characteristics of their equipment. Composite light-time corrected lightcurves are displayed in Fig. 2

Two events occur over the course of an orbit. The deeper event corresponds to the inferior event, whenever the largest component (called the primary for convenience) is being transited by the smaller one (secondary), whereas the shallower event corresponds to the superior event corresponding to the reverse configuration. Near opposition, the amplitudes of both inferior and superior events are quite similar and equal to 0.75 mag. On the other hand, pre-opposition and post-opposition lightcurves show significant changes in depth of the events, up to 0.85 mag for the inferior events and 0.8 for the superior events. These changes are due to mutual shadowing (eclipse event) which plays a major role in the morphology of mutual event ligthcurves. The depth of the events deepens and the lightcurves become also asymmetric compared to the no-shadow case (pure occultation event). The events become shallower as opposition is approached. The present observations, spanning an interval of 3 months, suffice to secure the reliability and the persistence of small photometric anomalies on the wings of the lightcurves. Thanks to the diversity of viewing and aspect conditions, information can be gleaned from these “anomalous” smooth photometric structures.

### 2.2 Updating the physical solution of the double system

Before addressing the possible causes of anomalies, the parameters of the system are slightly revised in the framework of the model proposed in Descamps et al. (2007). In the present work we used the Hapke bidirectional reflectance function (Hapke, 1981, 1984, 1986) to compute the amount of sunlight reflected from Antiope over the visible and illuminated surfaces. The bidirectional reflectance equation of Hapke describes the intensity of scattered light relative to the incident flux as a function of incidence angle, emission angle and phase angle. The Hapke law accounts for the opposition surge (at $\alpha<10°$) and adequately describes the disk-integrated brightness behavior as a function of phase angle. The Hapke photometric theory utilizes five parameters: $\tilde{\varpi}_0$ the single scattering albedo, $\bar{\theta}$ the macroscopic roughness, $g$ the Henyey-Greenstein asymmetry parameter, $B_0$ the amplitude of the opposition surge, and $h$ a regolith compaction parameter which characterizes the width of the opposition surge in terms of soil structure (porosity, compaction with depth). The main advantages of using the Hapke law instead of the empirical Minnaert law, used in our previous study, is that on the one hand the theory includes the effect of limb-darkening and on the other hand it is able to allow for the effects of a nonzero phase angle.

Hapke parameters can be initialized from the conversion equations proposed by Verbiscer and Veverka

(1995) which express each Hapke parameter as a function of the slope parameter *G* and the geometric albedo $p_v$. As we have to deal with relative photometry, we did not fit the slope parameter G to our data. An average value of G=0.09, typical for C asteroids (Harris and Young, 1988), $p_v$ = 0.0603 (inferred from IRAS observations, Davis and Neese, 2002) and a typical macroscopic roughness $\bar{\theta}$ = 20° (Helfenstein and Veverka, 1989) were chosen as input parameters. We derive the following parameters: $\tilde{\omega}_0$ = 0.064, h = 0.051, $B_0$ = 2.378 and g = -0.276. In relative photometry, the only relevant parameters are (h, g, $B_0$). For another G value of 0.15, (h,g) do not significantly change and the only slight variation is noted for the width of the opposition effect, $B_0$ which reduces to 2.0 instead of 2.4. These Hapke parameters values are very close to the ones derived for Phobos, satellite of Mars, which could be a carbonaceous C-type captured asteroid (Simonelli et al., 1998, Bibring et al., 1989).

Given this set of parameters, we carried out a new fit of the whole available lightcurves collected since 1996. The solar phase angle range is from -15° to +15°. A few near-opposition observations have been made around $5^{th}$ of January 2008 in order to further constrain the assumed Hapke parameters h and $B_0$. This phase angle coverage is sufficient to constrain the backscattering behavior (g parameter) and the opposition surge parameters (h, $B_0$). Eventually, we derived the same physical solution than in our previous study (Descamps et al., 2007). In other words, the Roche ellipsoidal solutions are kept identical (Table 3). Only, the pole position was refined to $\lambda_0$ = 199.5±0.5° and $\beta_0$ = 39.8±5° in J2000 ecliptic coordinates Thereby, we successfully and attractively substituted the Hapke law for the Minnaert law, used in our previous work. This can be taken as an a posteriori validation of the assumed Hapke photometric parameters, enabling us to now adequately reproduce the observed lightcurves at any phase angle.

### 3. Possible origins of the photometric anomalies

Two hypotheses may be invoked for the anomalies, but only one convincingly accounts for the observations. Photometric asymmetries could be caused by strength-supported topography or by the presence of albedo markings on the surface, perhaps analogous to those already mapped on the dwarf planet 134340 Pluto (Young et al., 2001) from mutual events with its locked satellite, Charon. We discuss these two possibilities and show that a unique large-scale non-convexity feature seems to be the most reliable explanation accounting for the photometric anomalies.

### 3.1 Albedo variegation

The first intuitive speculation consists of considering a dark spot, fainter than the rest of the surface. The most notable case of albedo variegation on a minor planet is displayed by 4 Vesta, which apparently has one bright and one dark hemisphere (Blanco and Catalano, 1979, Binzel et al., 1997, Li et al., 2008). However, albedo variegations frequently produce smaller brightness changes than the shaped–induced ones (Kaasalainen et al., 2002). In Fig.3, we investigated the photometric impact over the rotational lightcurve caused by an albedo pattern on the surface one of either component. For the sake of simplicity, this pattern is treated as a circular spot. We have assumed throughout that both components have uniform and identical albedo, apart from this small spot. The spot has four parameters: the relative albedo with respect to the rest of the surface, the latitude and longitude of its center and its angular radius. The synthetic lightcurves, generated with a circular spot and displayed in Fig.3, show that we cannot reproduce observed anomalies at all. Whenever the spot becomes visible to the observer, the lightcurve tends to be flattened. Furthermore, this effect appears only once in the lightcurve whereas anomalies occur twice per rotation.

### 3.2 The giant crater hypothesis

A second explanation for this repetitive anomaly could be due to the presence of a large-scale geological feature, such as a giant crater. Specifically the term « giant crater » is used to denote craters with a diameter $D_c$ comparable to the mean radius $R$ of the body such as $D_c > 0.75R$. Furthermore, this hypothesis is motivated by the existence of such asteroids on which giant craters have been already observed. For example, the Near Earth Asteroid Rendezvous (NEAR) spacecraft revealed at least five craters of diameter 19 to 33 km on the C-type asteroid 253 Mathilde, comparable to its 26.5 km mean radius. The giant craters observed on Mathilde are thought to be formed by compaction of material after an impact, due to the extreme macro-porosity of its interior. They have, therefore, a « simple » bowl-shaped morphology. Accordingly, our model of a large-scale depression on the surface of a component consists of a hollow hemispherical portion of radius $R_e$, whose the center is located above the surface at an altitude $h$. Assume a crater diameter $D_c$ and its depth $d = D_c/t$, where $t$ is a constant characterizing the diameter/depth ratio of the crater. The diameter/depth ratio, t, can be derived from the excavating sphere parameters by the simple following formula:

$$\frac{h}{R_e} = \frac{t^2 - 4}{t^2 + 4} \qquad (1)$$

A realistic range of values for the ratio $t$ is given by $t \geq 2$. For instance, photoclinometry applied to fresh craters on Ida (Sullivan et al., 1996) indicates a ratio t~6.5. Craters on 433 Eros show similar ratios (Veverka et al., 2000). Fig. 4 shows the photometric effect of each of these parameters, including the location in longitude/latitude over the body of the crater center. It quickly appears that only a large crater with a significant depth is able to satisfactorily account for anomalies.

A trial and error fit of the large-scale feature parameters was performed with all available observations published in literature using a goodness-of-fit criterion, taken as the averaged differences between simulations and observations. The criterion Θ is defined as:

$$\theta\ (mag) = \sqrt{\frac{\sum_{i=1}^{n} (O_i - C_i)^2}{n}} \qquad (2)$$

Where n is the data number, and $O_i$ and $C_i$ are observed and calculated magnitudes. Fig. 2 shows the lightcurve of the best-fit model which overall agree to better than 0.02 mag with the observations whereas the pure Roche model gives a value of Θ ~ 0.03 mag, vindicating the large-scale feature approach. Fig. 5 show the model lightcurves compared with past observed lightcurves. The first complete lightcurve of 90 Antiope, which revealed the large amplitude (0.7 mag) and was recorded in December 1996 by (Hansen et al., 19997), can be accurately fit with our new shape model including the large crater. A shallow (Δm=0.15) limb-grazing event lightcurve observed in December 2001, when the system was substantially tilted on the line of sight (Michalowski, 2002), can be also well reproduced with the new model. Residual discrepancies of ~0.001 mag, seen in December 2001, comes from the fact that surfaces are quite obviously not perfectly ellipsoidal and likely exhibit some other small-to-medium scale topological features. It should be kept in mind that the model is a first-order model, including the main physical photometric effects, such as the limb-darkening, the mutual shadowing controlled by the solar phase angle and the large-scale shape effects. The adopted best-fit solution for the crater gives a diameter/depth ratio, t = 3, with a diameter of ~68 km, i.e., a fraction of ~0.7 of the average diameter of the components of Antiope. However, it is worth pointing out that it is not possible to settle which component hosts the crater and we have arbitrarily decided to assign it to the largest one. The topographic location of its center lies on the trailing side of the body at 145±5° in longitude and 40±5° in latitude. The bulk density inferred from the pure Roche solution is 1.24±0.04 g/cm$^3$. After including the crater, the total volume is reduced by about 3%, entailing an equivalent increasing of the bulk density which is then equal to 1.28±0.04 g/cm$^3$. Fig. 6 shows a picture of the Antiope system, generated from the present model, as viewed on November 3 2007 at 10:25 UTC, just at the onset of a mutual eclipse of the secondary by the primary.

## 4. Discussion

### 4.1 Conditions for the existence of a giant impact crater on Antiope

Antiope is a member of the Themis family, one of the largest currently known asteroid families with about 550 identified members (Zappalà et al., 1995) with a size limit of completeness of ~20 km. According to Migliorini et al. (1995), the expected number of interlopers is relatively low, even if one or two can be expected at sizes as large as 40 km. The age of the Themis family was estimated between a few hundred million years and ~2.3 Byr (Marzari et al., 1995), and recently refined to about 2.5 ± 1.0 Byr (Nesvorny et al., 2005). The Themis family is supposed to have been formed by the disruption of a large 450-km size parent body whose the only 15% probability makes this event as probably unique (Marzari et al., 1995). Antiope is therefore thought to be the outcome of the catastrophic disruption of one of the largest asteroids in the main-belt that broke it into multiple pieces. We can wonder under which conditions a 100-km proto-Antiope body can survive high-velocity impacts without being disrupted and fully dispersed.

The physical and morphological characteristics of the postulated crater are quite similar in proportion to the giant crater seen on 253 Mathilde. Its depth is only one third of its diameter (t = 3). Various laboratory experiments on the depth/diameter ratios, for high-velocity impacts against the projectile-target density ratio, have shown that diameter/depth ratios lower than about 3 is highly suggestive of a low-density and porous surface structure of the target (Housen et al., 1999, Housen and Voss, 2001,

Nakamura, 2002 and references therein). Furthermore, an outstanding characteristic of 90 Antiope is precisely to be a highly porous C-type asteroid with a macroscopic porosity of about 50%, derived from its low density (Descamps et al., 2007). The interior can therefore be considered as a loosely bound pile of material, usually referred to as "rubble pile". An impact on such a porous object will compact material on a local area, having almost no effect on the internal structure and shape of the asteroid. The shock wave, produced by the impact, dies out quickly, and a large bowl-shaped crater can be formed (Asphaug et al. 1998). Love et al. (1993) have shown that it requires more projectile energy to produce the same cratering effect in a porous target than in a non-porous target. The C-type main-belt asteroid 253 Mathilde, imaged by Near Earth Asteroids Rendezvous (NEAR) spacecraft, is a peculiar example. The images revealed the presence of several giant craters larger than conventionally accepted crater size limit for disruption (Chapman et al., 1998), confirming the existence of this phenomenon. More recently, experimental studies on crater morphology for hypervelocity impacts on highly porous targets (Giacomuzzo et al., 2007) have led to the confirmation of the following relation, empirically obtained by Kadono et al. (1999), for crater diameter ($D_{max}$) to projectile diameter ($D_p$) ratio as a function of impact velocity:

$$\frac{D_{\max}}{D_p} = 10^{-0.07\pm 0.05} V_{imp}^{1.3\pm 0.1} \quad (3)$$

They proposed also a relation for the penetration depth $p_{max}$ as a function of the projectile density ($\rho_p$) to the target density ($\rho_t$) ratio:

$$\frac{p_{\max}}{D_p} = 10^{0.33\pm 0.31} \left( \frac{\rho_p}{\rho_t} \right)^{1.07\pm 0.17} \quad (4)$$

In this formalism, impacts are not oblique so that the projectile density obtained should be considered as a lower limit density. After working out equations [2] and [3] we derive, against the impact velocity, the mass ratio $m_p/M_T$ required to produce an impact crater of diameter $D_{max}$ :

$$\frac{m_p}{M_T} = \frac{\rho_p}{\rho_T}\left(\frac{D_p}{D_T}\right)^3 = 0.6857t^{-1/1.07}\left(\frac{D_{\max}}{D_T}\right)^3 V_{imp}^{-2.687} \qquad (5)$$

We have plotted the mass ratio versus the impact velocity in Fig.7 for the nominal values of the crater characterized in Section 3.2 ($D_{max}$=68 km, $p_{max}$ = 23 km and $D_T$=100 km). Limiting curves stem from the range of variation of the exponents in the equation [3,4]. Besides, the distribution of relative velocities between members of the Themis family is nearly uniform between 1 and 5 km/s, with a mean velocity of 3.36 km/s (Bottke et al., 2002). Accordingly, if we assume that the impactor is a family member, we can infer from Fig. 7, for this mean relative velocity, a mass ratio q = 0.005±0.004 or a diameter of the impactor of 17 +4/-12 km.

### 4.2 Implication for the possible origin of the double system of Antiope

The presence of this large geological feature could be linked with the binary nature of the asteroid. The most common explanation of how objects acquire moons involves the collision of two bodies, after which some of the collisional debris reaccrete and end up in orbit around the larger body (Durda et al., 2004). The system of Antiope could be the aftermath of the collision between two large fragments. However, the collision scenario cannot produce two same-sized bodies orbiting each other (Weidenschilling et al., 2001). Descamps et al. (2007) suggested another mechanism through which Antiope could originate from a single, loosely bound and fast-rotating body. An oblique impact - perhaps due to the encounter with another fragment of the Themis family - could have imparted even more spin to the proto-Antiope; if the asteroid is spun up fast enough, centrifugal force can have overcome the force of gravity. The splitting of this 100-km parent body in two components follows a sequence analogous to the dumb-bell sequence for bodies with hydrostatic equilibrium figures (Eriguchi et al., 1982). The hypothesis that the two bodies may have begun as one parent body is supported by the fact that assumptions such as similar reflectivity properties and similar bulk density for either component holds very well to account for the overall observations. Furthermore, the trailing side location of the crater substantiates the idea of a rear impact which could have sped up a proto-Antiope in the same sense as its actual rotation.

The angular momentum of the projectile is $L_p = m_p V_{imp} R_T / \sqrt{2}$ (Weidenschilling, 1989) while the threshold of rotational fission is reached at $L_{th} \sim 0.5\sqrt{G M_T^3 R_T}$ , where G is the gravitational constant. This threshold corresponds to the transition between the equilibrium dumb-bell sequence and the double sequence of hydrostatic spinning fluid masses (Descamps and Marchis, 2008). Accordingly, fission requires the following condition to be fulfilled:

$$\frac{L_p}{L_{th}} \approx 1 \qquad (6)$$

which can be rewritten as follows:

$$\frac{m_p}{M_T} \approx 0.5 \frac{V_e}{V_{imp}} \qquad (7)$$

Where $V_e = \sqrt{8/3\pi \rho G R_T}$ is the escape velocity from the parent proto-Antiope body. For a proto-Antiope having a radius $R_T$=50 km we have $V_e$= 42 m/s. After having superimposed in Fig. 7 the condition for fission [7], it is straightforward to infer the allowed range of impact velocity required for fission which is about 1.0-4.2 km/s as well as the minimum mass ratio of 0.005 at 4.2 km/s or 17 km in diameter. We see that as soon as a projectile, issued from the Themis family, has a size on the order of 20-30 km, it will be able to make the target fission provided that its velocity is adequate. Once formed, the system will evolve by tidal interaction over very short timescale. This timescale can be computed by the equations described by Weidenschilling (1989). In Fig. 8 we have plotted tidal evolution time scales as a function of the relative separation, and mass ratio (Weidenschilling et al., 1989), we may put corresponding values derived for Antiope, $a/R_p=3.8$ and $q=0.95$. We have adopted $\mu Q \approx 10^{12}$dynes/cm$^2$. The specific energy dissipation function Q is generally ~100. Moderately fractured carbonaceous asteroids (such as Phobos) have a coefficient of rigidity $\mu \sim 10^{10}$dynes/cm$^2$. We can bind the evolution time, assuming initial $a/R_p=1$, to ~10 000 years.

### 4.3 Frequency of an intra-family impact

The frequency of a collision of a 100-km proto-Antiope with another body of the Themis family, as large as 17 km, over the age of the family should be addressed in order to assess whether or not this hypothesis may be considered as realistic. We know that there are about 250 potential projectiles with a diameter greater than 17 km in the Themis family. Using the mean intrinsic collision probability $P_i=10.1\text{x}10^{-18}$ km$^{-2}$ year$^{-1}$ per crossing pair of bodies of the Themis family (Bottke et al., 2002), the collision frequency[1] on a ~100 km target of such a projectile is of order $0.2.10^{-9}$/y. Therefore, with a substantial 50 % probability of having occurred over the age of the family (2.5 $10^9$yr), such an intra-family impact is highly probable. We can note that this probability is only of 14 % for a projectile with a diameter greater than 27 km (~60 bodies in Themis family).

[1] The Themis family contains 7 bodies of diameter ~100km and a population of ~250 members with $D > D_{mini}=17$km. The mean collision frequency is then computed by multiplying the intrisinc collision probability, $P_i$, by the total number of pairs of bodies and by the geometrical cross section of the bodies, $\pi(R_t+R_{mini})^2$. We derive the collision rate with a body of 17 km in diameter, $f_i = (7\text{x}250)\ \pi\ (50+8.5)^2 10.1\text{x}10^{-18} = 0.2\ 10^{-9}\ \text{yr}^{-1}$

## 5. Conclusion

The observed lightcurves of 90 Antiope, during mutual event seasons, are consistent with system made of hydrostatic equilibrium ellipsoids, including on one of the components a bowl-shaped crater. A large dark spot, having an albedo very different from the surroundings, has been ruled out by our analysis. We propose that the Antiope system originates through the fissioning of a 100-km sized parent body consecutive to a violent impact with another smaller member of the Themis family. This projectile, about 20 km in diameter, would be responsible of the prominent non-convex feature postulated in the present work. Thanks to its high porosity, a proto-Antiope can survive such a catastrophic collision which otherwise would have destroyed any coherent or monolithic body. This event appears to have a substantial 50 % probability over the age of the Themis family.

**Acknowledgements**

F.M. work is supported by the National Aeronautics and Space Administration issue through the Science Mission Directorate Research and Analysis Programs number NNX07AP70G. The work of T.M. was supported by the Polish Ministry of Science and Higher Education - grant N N203 302535. Data from SF, MF and JPS have been obtained with the Pic du Midi Observatory 0.6 m telescope, a facility operated by Observatoire Midi-Pyrénées (France) and Association T60, an amateur association.

**Table 1:**

List of the observers, their facilities and filter(s) used for the observations

| Observers | Observatory | Aperture (m) | filter |
|---|---|---|---|
| Colas, F. | Pic du Midi Observatory<br>IAU code #586<br>5°27'21”E<br>43°18'38”N | 1.00 | Large filter, V+R |
| Descamps, P., J. Berthier, F. Vachier | Haute-Provence Observatory<br>IAU code #911 | 1.20 | R |
| Fauvaud, S., Sareyan J.-P., Fauvaud, M. | Pic du Midi Observatory<br>IAU code #586<br>5°27'21”E<br>43°18'38”N | 0.60 | R |
| Marchis F. | Lick Observatory<br>IAU code #662 | 1.00 | B |
| Pilcher, F. | Organ Mesa Observatory | 0.35 | Clear |
| Pollock, J. | Appalachian State University,<br>Rankin Science Observatory<br>81°40’ 54”W<br>36°12’ 50”N | 0.40 | R |
| | PROMPT, CTIO<br>Cerro-Tololo<br>70°48’14”W<br>30°10’8”S<br>IAU code #807 | 0.41 | R |
| Klinglesmith, D.A. | Magadalena Ridge Observatory<br>IAU code #H01 | 0.35 | Bessel R |
| Wiggins P. | 112°18'E<br>40°38'N<br>IAU code #718 | 0.35 | Clear |

**Table 2:** Orbital elements and characteristics of 90 Antiope system. Elements are expressed with respect to the mean Equator and Equinox of J2000.0 (Descamps et al., 2007)

| | S/2001 (90) 1 |
|---|---|
| Period (h) | 16.5051±0.0001 |
| Semi-major axis (km) | 171±1 |
| Orbit Pole solution in ECJ2000 (degrees) λ(longitude), β (latitude) | $\lambda$=200+/-2°, $\beta$=38+/-2° |
| Inclination (degrees) | 63.7+/-2° |
| Ascending node (degrees) | 303.1 +/-2° |
| Mass (kg) | $8.3 \pm 0.2 \times 10^{17}$ |

**Table 3**: Best Roche ellipsoidal solution of the system of 90 Antiope. The crater reduces the total volume by about 3% which in turn increases the bulk density to 1.28 ±0.04 $gcm^{-3}$.

| Component A | | | Component B | | | Separation | Size ratio | Density |
|---|---|---|---|---|---|---|---|---|
| a | b | c | a' | b' | c' | | | |
| km | km | km | km | km | km | km | | $g/cm^3$ |
| 46.5 | 43.5 | 41.8 | 44.7 | 41.4 | 39.8 | 171 | 0.95±0.01 | 1.24±0.04 |

**Fig. 1 -** Composite ligthcurve of Antiope taken at SAAO on June 2005 (Descamps et al., 2007). The dashed curve results from a pure Roche solution made of two ellipsoids. Note the shallow concavities just after the eclipse egresses, centered at a rotation phase of 0.35 and 0.85. These anomalies can depart from the perfect Roche model by an amount of 0.04 mag. The solid line corresponds to the best-fit model curve obtained from addition of a large-scale feature described in Section 3.

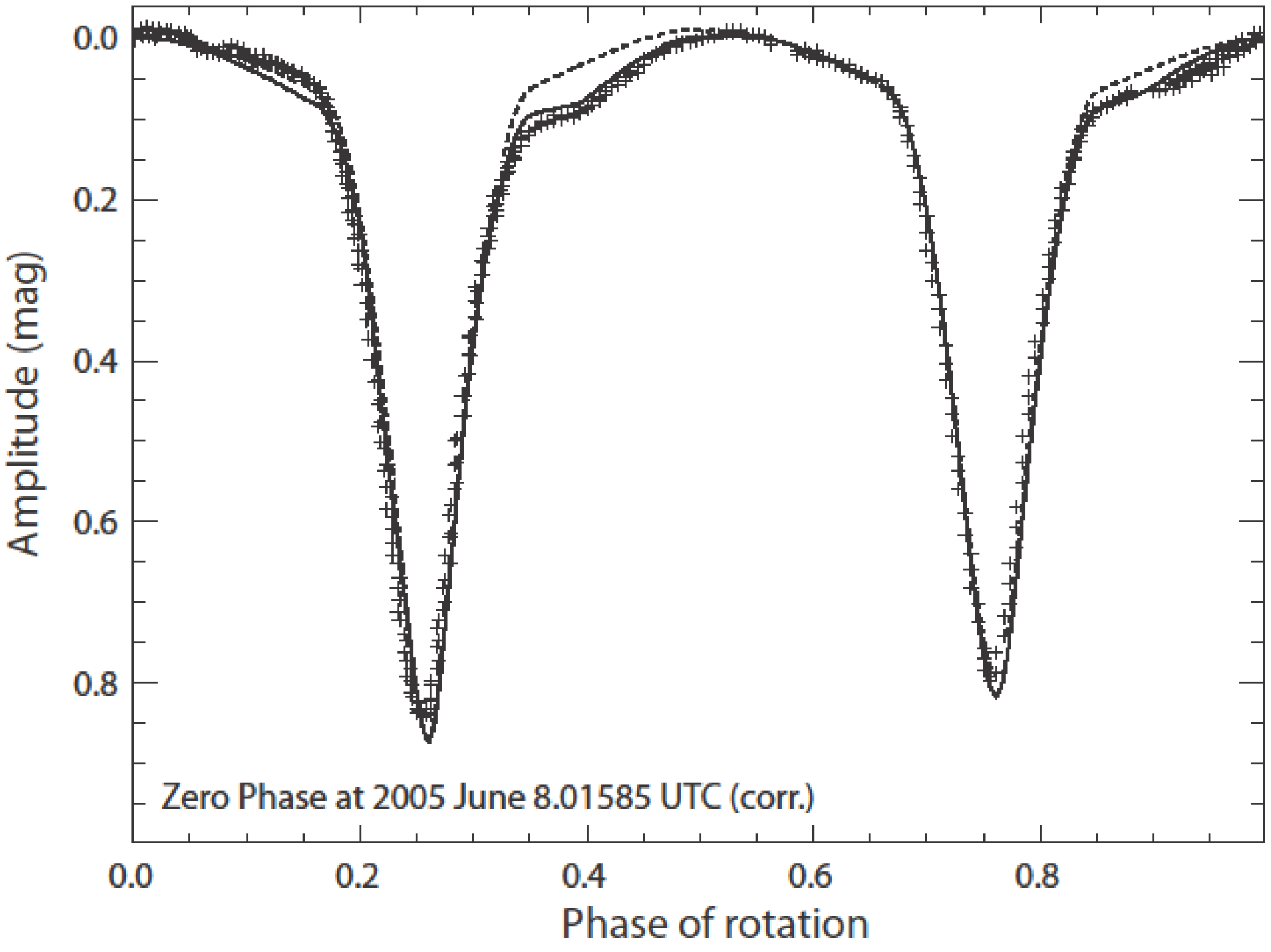

**Fig. 2** - Rotational lightcurves, corrected for light time, collected during the 2007-2008 campaign. The synthesized lightcurves (solid line), endowed by the Roche solution derived by Descamps et al. (2007) including a giant crater on either component, are superimposed to the observations (symbols).

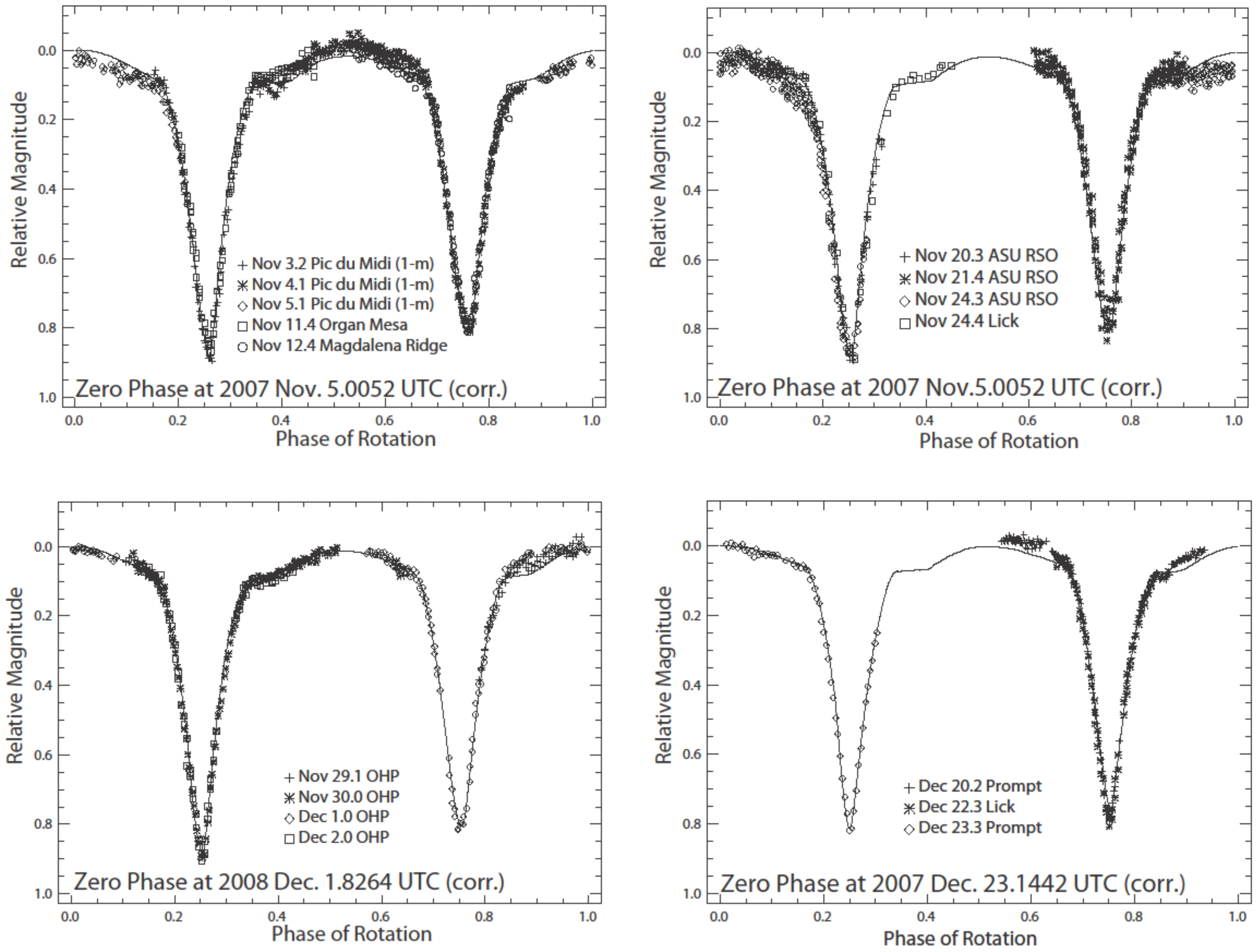

**Fig. 2** – continued.

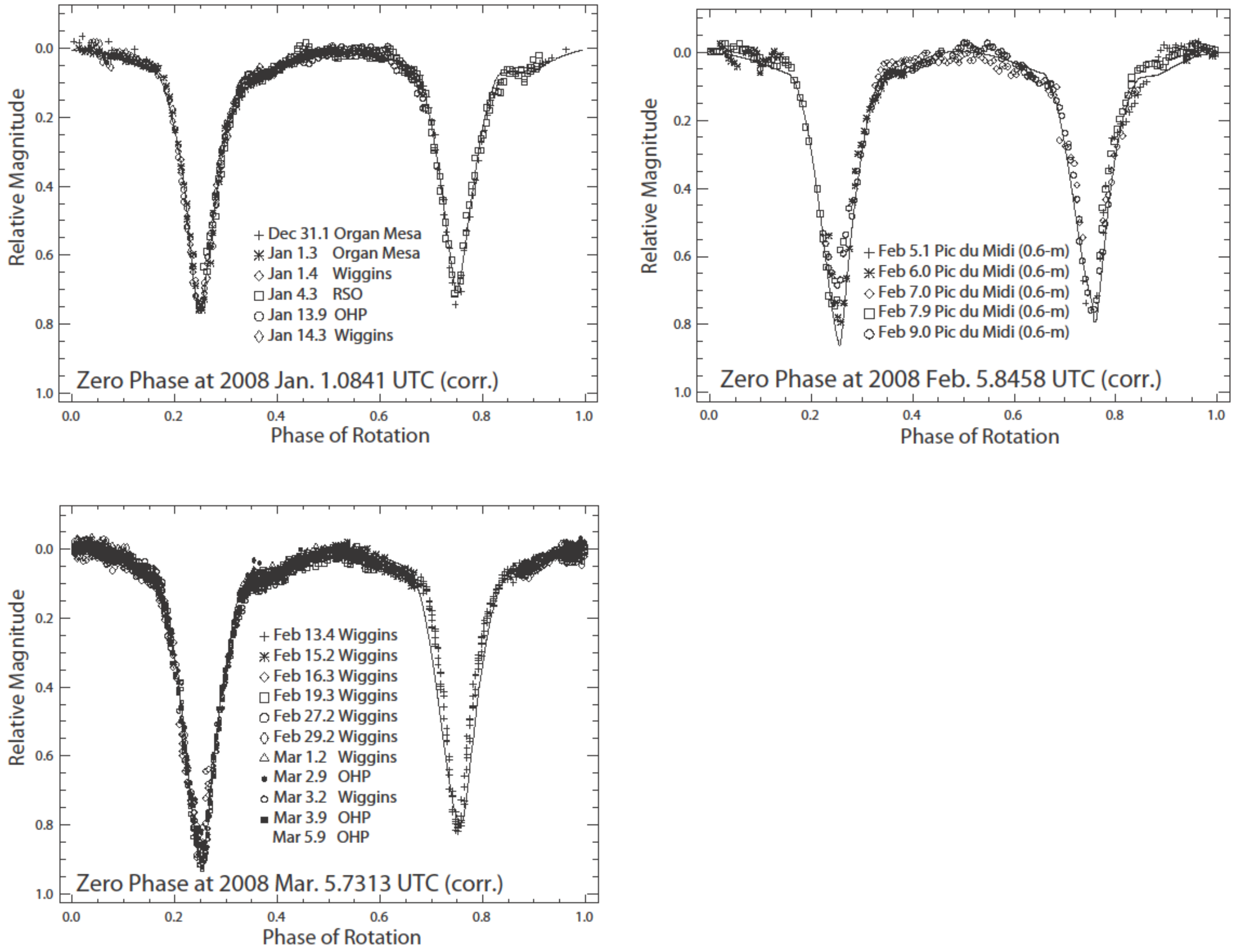

**Fig. 3 -** Photometric effect of an albedo patch on the surface. These lightcurves were generated adding a circular pattern on the Roche ellipsoid model with different characteristics: the relative albedo of the dark patch with respect to the rest of the surface, the angular radius of the crater, the location of the crater over the surface in longitude and latitude.

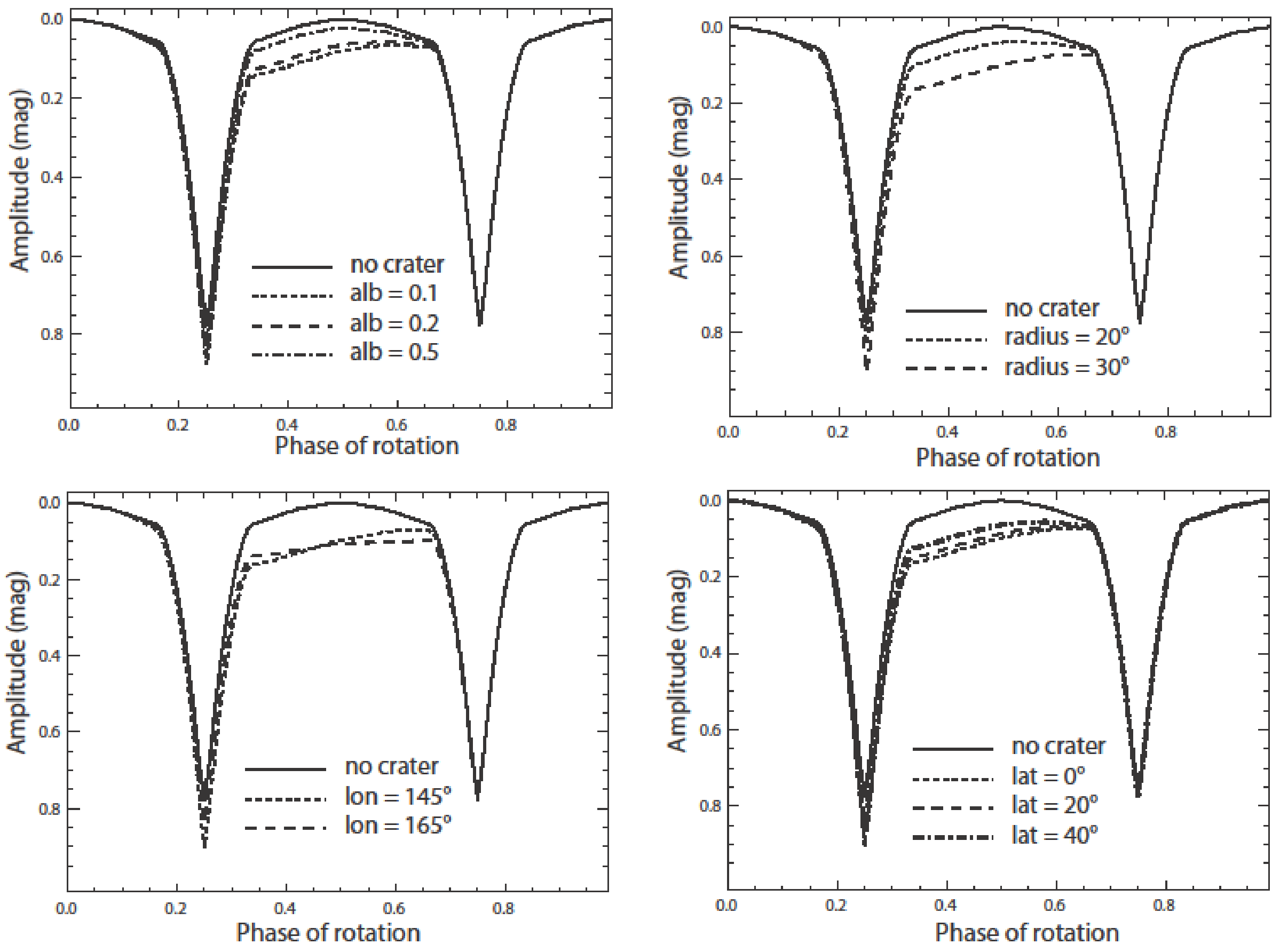

**Fig. 4 -** Photometric effect of crater parameters. These lightcurves were generated adding a crater on the Roche ellipsoid model with different characteristics: the dimensionless depth parameter t, the ratio of the crater radius $R_c$ to the radius of the body R, the location of the crater over the surface in longitude (L) and latitude (l). The curves in plain line correspond to the reference set of values: t = 3, $R_c$ /R = 0.7, L = 145° and l = 40°. Each panel displays variations of a single parameter while holding the others at their nominal values.

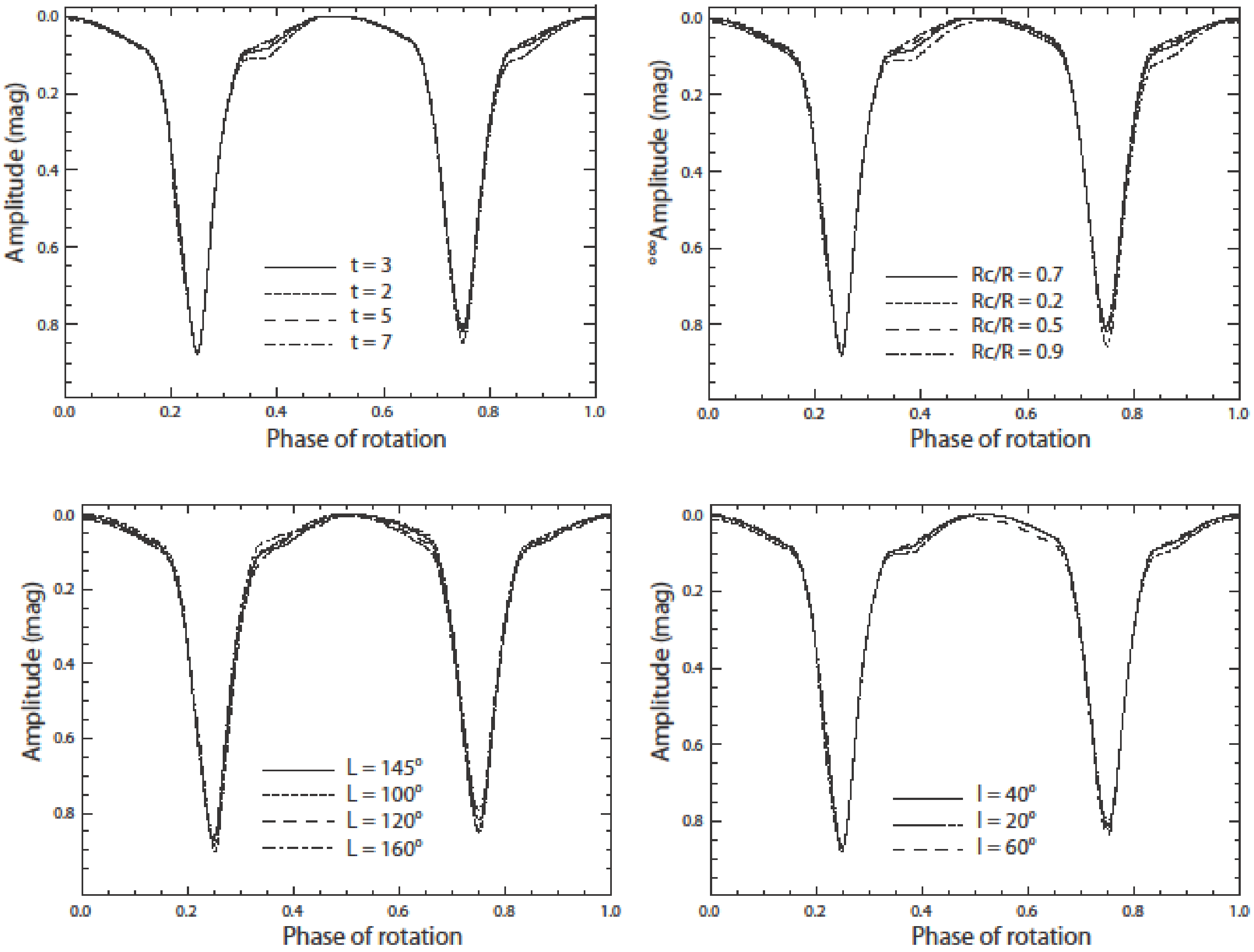

**Fig. 5** - Past lightcurves recorded in 1996 and 2001. The first lightcurve of Antiope, made in 1996, exhibiting a substantial amplitude of 0.7 mag (Hansen et al., 1997). In 2001, the system of Antiope was viewed under an almost pole-on aspect giving rise to grazing events (Michalowski et al., 2002). The Roche model without crater has been overplotted in dotted line. Our updated model, including the giant crater, is superimposed to the observed lightcurves. Residual discrepancies of ~0.001 mag, seen in December 2001, comes from the fact that surfaces are not perfectly ellipsoidal and likely exhibit some other topological features at lower scales.

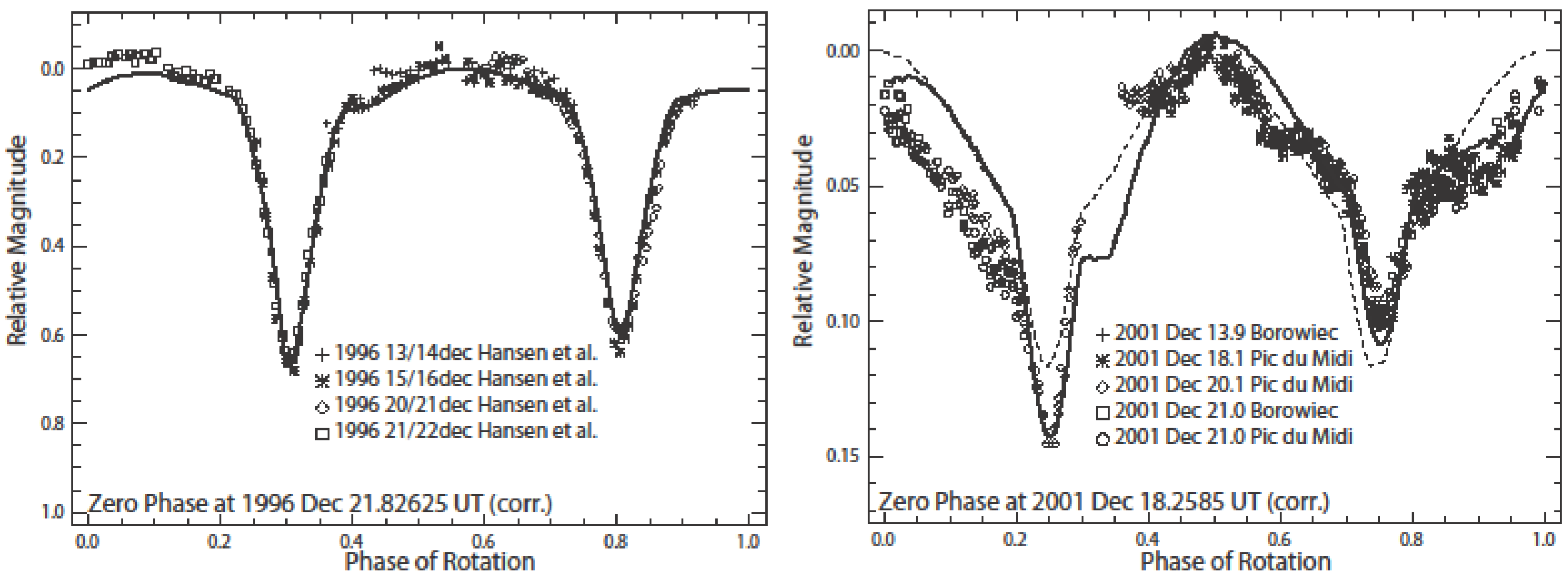

**Fig. 6 -** Picture of the double system of 90 Antiope showing the modeled crater with a diameter of 68 km and depth of a third its diameter (t=3). The crater lies on the trailing side of the component at 145° in longitude and 40° in latitude. The system is represented on November 3 2007 at 10:25 UTC just at the onset of a mutual eclipse of the secondary by the primary. The relative path of the secondary is also plotted. North is up and East is on the left. The spatial resolution on each component is 1°. Rendering is made from Hapke law. The sub-Earth point has a longitude of 115.7° and a latitude of -1.7°. Position angle of the North pole is 58.9°. The phase angle is 15.4°.

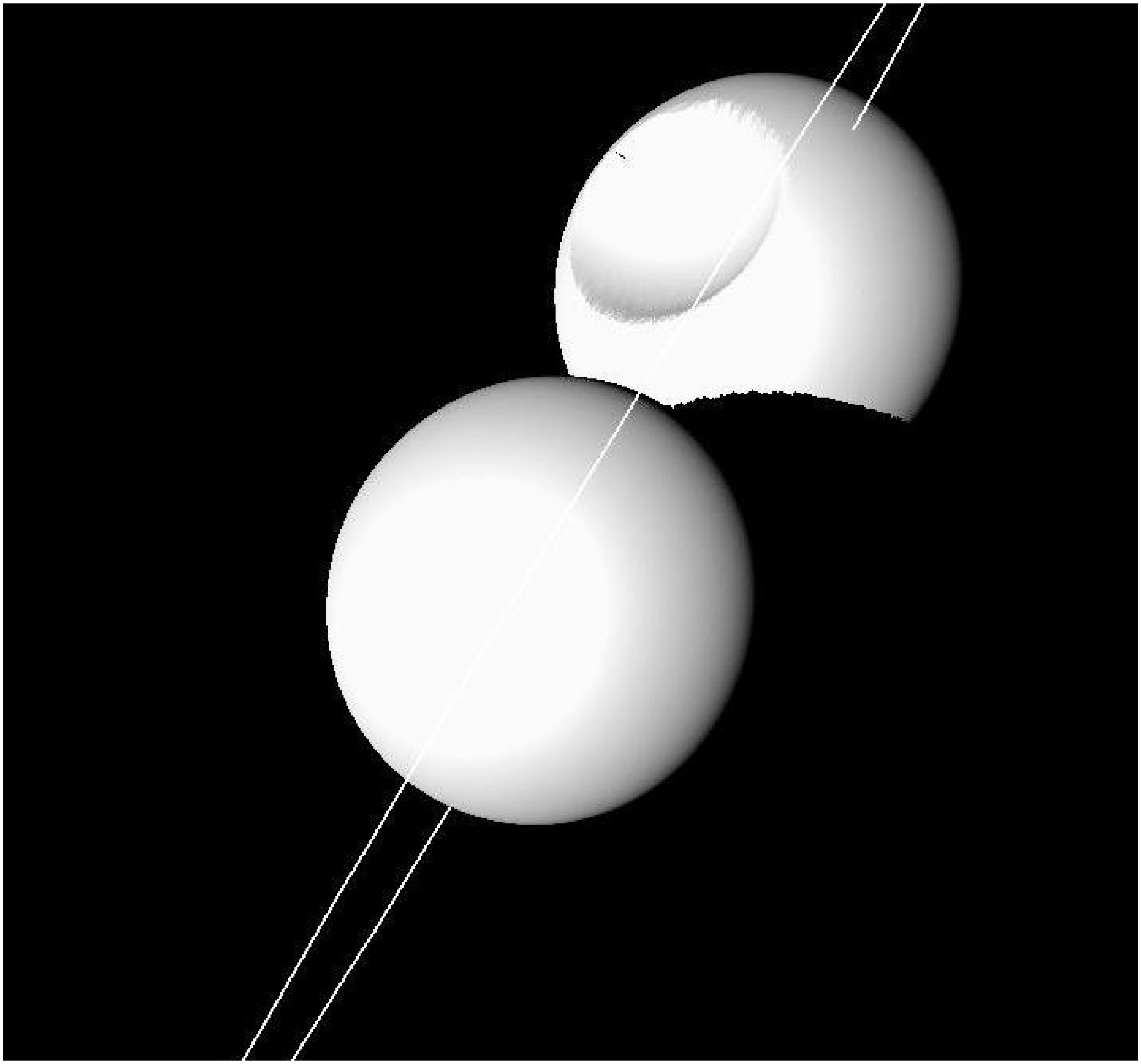

**Figure 7:**
Projectile to target mass ratio versus the impact velocity for a 100-km target. The admissible zone for the projectile parameters (mass ratio and impact velocity) able to produce the postulated crater is bound by the dashed curves (given by equation [5]). For all values of the projectile parameters located within this admissible zone and above the limit of fission (plain line curve plotted from equation [7]), the target will be spun up to fission by the impact. The velocity range of Themis family members is indicated as well. We can see that an intra-family projectile ranging from 17 to 27 km in diameter can readily bring the target (potentially the parent body of Antiope) to fissioning and be responsible of the postulated crater.

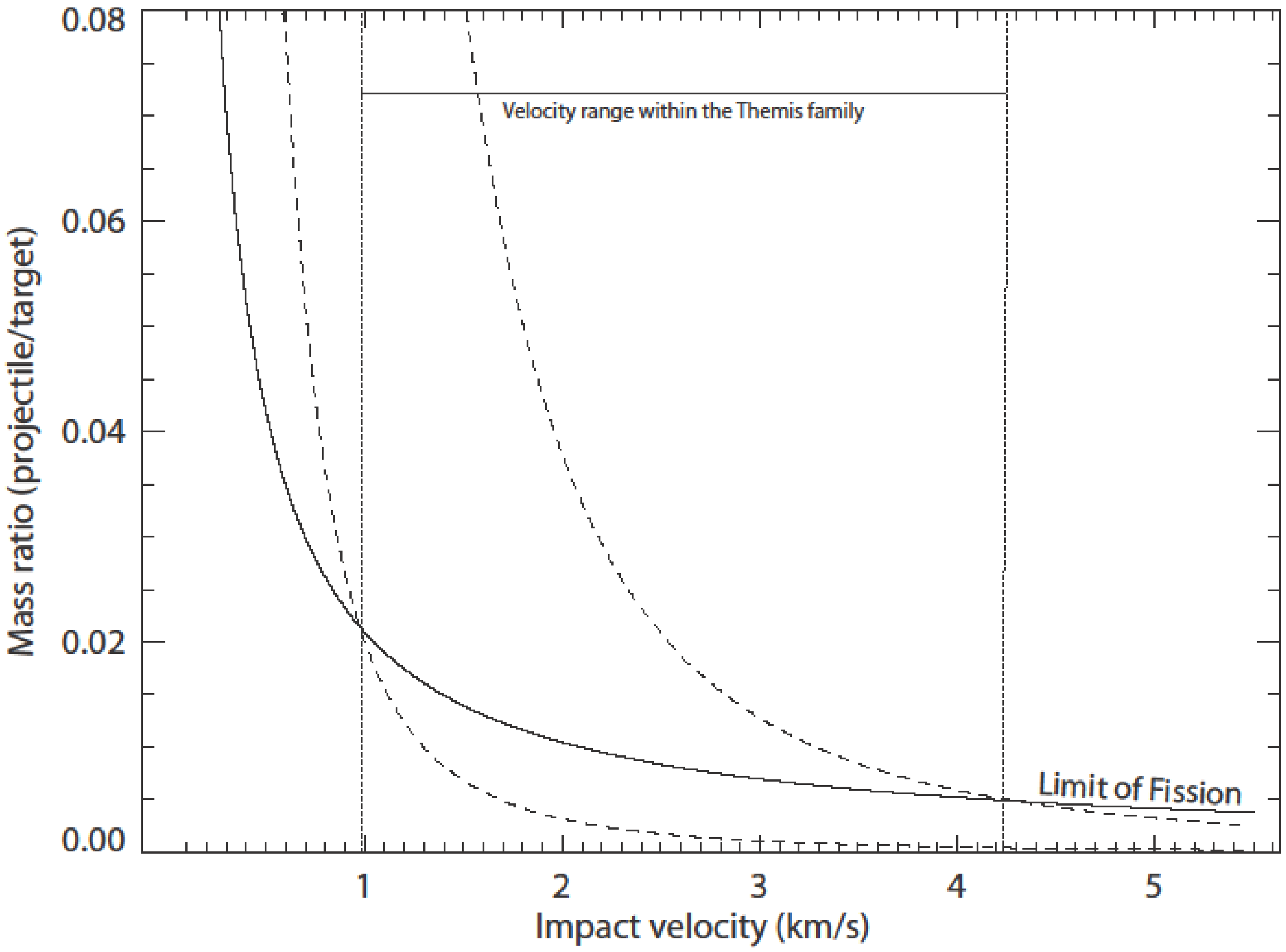

**Figure 8:**
Diagram of the relative separation a/Rp against the mass ratio q after Weidenschilling et al. (1989). Curves of isotimescale are plotted. They assume initial a/R=1, $\mu Q \approx 10^{12}$dynes/cm$^2$, $\rho$=1.28g/cm$^3$, Rp=45km and q = 0.95 which are the physical characteristics of Antiope derived in the present work. Binaries to the left of the curve labelled synchronous stability cannot maintain spin-orbit synchronism. The location of Antiope in this diagram shows the quick tidal timescale necessary to evolve towards synchronisation, ~10000 years.

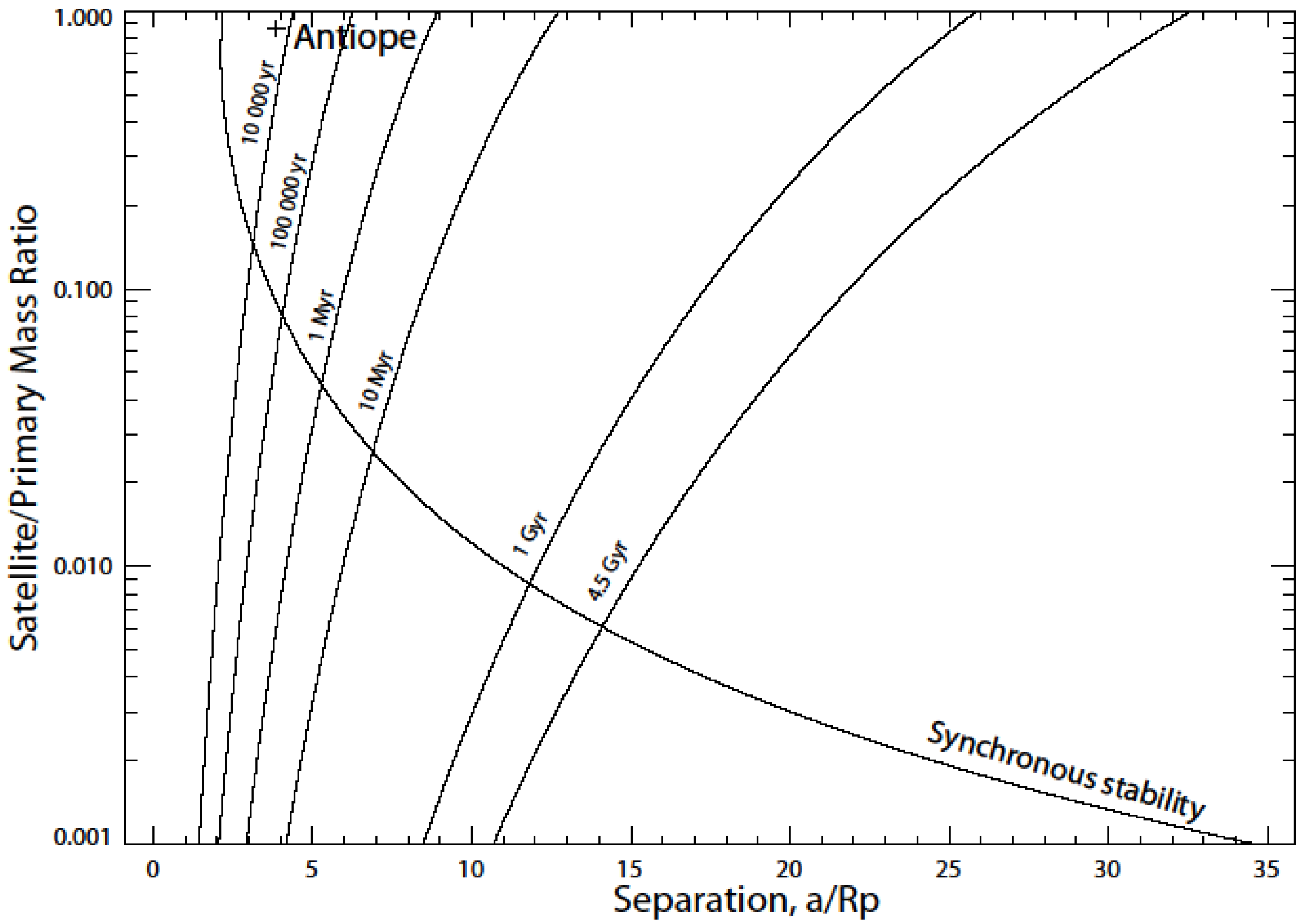